\documentclass[12pt,preprint]{aastex}

\shorttitle{Habitable Planet Formation in Binary-Planetary Systems}
\shortauthors{Haghighipour \& Raymond}

\begin{document}

\title{Habitable Planet Formation in Binary-Planetary Systems}

\author{Nader Haghighipour}
\affil{Institute for Astronomy and NASA Astrobiology Institute,\\
University of Hawaii-Manoa, 2680 Woodlawn Drive,
Honolulu, HI 96822}
\email{nader@ifa.hawaii.edu}
\and
\author{Sean N. Raymond}
\affil{Center for Astrophysics and Space Astronomy, and
Center for Astrobiology,\\ 
University of Colorado, Boulder, CO 80309\\}
\email{raymond@lasp.colorado.edu}

\begin{abstract}

Recent radial velocity observations have indicated that
Jovian-type planets can exist in moderately close binary 
star systems. Numerical simulations of the dynamical 
stability of terrestrial-class planets in such environments
have shown that, in addition to their giant planets, these 
systems can also harbor Earth-like objects. In this paper, we
study the late stage of terrestrial planet formation in such 
binary-planetary systems, and present the results of the 
simulations of the formation of Earth-like bodies in their 
habitable zones. We consider a circumprimary disk of Moon- 
to Mars-sized objects and numerically integrate the orbits of 
these bodies at the presence of the Jovian-type planet of the 
system and for different values of the mass, semimajor axis, 
and orbital eccentricity of the secondary star. Results 
indicate that, Earth-like objects, with substantial amounts of 
water, can form in the habitable zone of the primary star. 
Simulations also indicate that, by transferring angular momentum 
from the secondary star to protoplanetary objects, the giant planet 
of the system plays a key role in the radial mixing of these 
bodies and the water contents of the final terrestrial planets.
We will discuss the results of our simulation and show that
the formation of habitable planets in binary-planetary systems 
is more probable in binaries with moderate to large perihelia.

\end{abstract}


\section{Introduction}

For the past several years, the formation of terrestrial planets 
in binary star systems has been the subject of research by many authors. 
\citet{Quintana02}, \citet{Quintana03}, and \citet{Lissauer04} 
studied the interactions of planetesimals and protoplanetary objects 
around the stars of $\alpha$ Centauri, and showed that, 
terrestrial-type planets can form around these stars when dynamical 
friction is included in numerical simulations. \citet{Barbieri02} 
and \citet{Turrini05,Turrini06} also studied terrestrial planet 
formation in this system and by considering gas-drag as the primary 
force for reducing planetesimals relative velocities, showed that, 
it is indeed possible to form terrestrial-class objects around the 
primary of $\alpha$ Centauri stars. In a recent article, 
\citet{Quintana07} have extended their simulations to wider binary 
systems, and identified regions of the parameter-space for which 
terrestrial planets can form around the stars of the binary.

The studies of terrestrial planet formation in dual-star systems,
as presented in the literature, share one common ground: The systems 
considered in these studies do not contain giant planets, and the 
formation of terrestrial planets has been simulated only at the 
presence of the two stars of the system. In this paper, we extend 
these studied to more complex environments and simulate the formation 
of terrestrial planets in binary star systems in which the primary 
star is host to a Jupiter-like planet. The purpose of our study is 
to understand how, in such binary-planetary systems, the dynamics 
of the stellar companion will affect the formation and the water 
contents of Earth-like objects in the habitable 
zone\footnote{The region around a star where a
terrestrial-class planet can maintain liquid water on its surface.} 
of the primary star.

The systems of our interest are moderately close ($\leq 40$AU)
binaries. Recent detections of Jovian-type planets in such environments 
(e.g.,  GJ 86, cf. Els et al. 2001, $\gamma$ Cephei, cf. Hatzes et al. 2003) 
have raised questions about the formation of these objects
and the possibility of the existence of smaller bodies in these systems.
In regard to the latter, simulations of the orbital dynamics of 
terrestrial planets in the $\gamma$ Cephei planetary system have
indicated that small objects can have long-term stable orbits in 
binary-planetary systems provided their orbits lie outside the influence 
zone\footnote{The influence zone of a planetary-sized object
with a mass $m_p$ is defined as the region between ${a_p}(1-{e_p})-3{R_H}$ 
and ${a_p}(1+{e_p})+3{R_H}$, where $a_p$ is the planet's 
semimajor axis, $e_p$ is its orbital eccentricity, and 
${R_H}={a_p}{({m_p}/3M)^{1/3}}$ represents its Hill radius. 
The quantity $M$ denotes the mass of the central star.} 
of the system's giant planet and are limited to the region between 
this object and its host star \citep{Hagh06}. In this study, we 
choose systems in which the primary star has a Jovian-type body 
in an orbit outside its habitable zone.

Unlike the majority of the binary-planetary systems that have 
so far been discovered\footnote{See \citet{Hagh06} for a complete 
and up-to-date list of binary-planetary systems.}, the relatively 
smaller separations of the binary components in our systems imply 
that the effect of the farther companion on the formation and 
dynamical evolution of planets around the primary star is not 
negligible. For instance, the perturbative effect of the secondary 
star can change the structure of the circumprimary disk and truncate 
it to a smaller size \citep{Artymowicz94}. The latter removes 
material that may be used in the formation of planets. For this 
reason, it was believed that circumstellar disks around the components
of a close binary may not be large and massive enough to begin
planet formation. However, observations by \citet{Math94}, 
\citet{Akeson98}, \citet{Rodriguez98}, and \citet{Math00}
have shown otherwise. These observations confirm the presence of
potentially planet-forming environments around the components of 
binary stars and imply that planet formation around a star of a 
binary may be as common as around a single star. In fact, the 
observations of two well-separated disks in the binary system of 
L1551 by \citet{Rodriguez98} indicate that, despite of disk truncation, 
it is still possible for the both components of a binary to
retain a relatively significant amount of their original 
circumstellar materials (0.03 to 0.06 solar-masses) in disks with
considerable radii ($\sim$ 10AU). The masses of these disks are 
comparable to the minimum solar-mass model of the primordial nebula 
of our solar system \citep{Stu77,Hayashi81}, implying that, planet 
formation in dual-star systems can begin and continue in the same 
fashion as around our Sun.

In this paper, we base our study on the latter consideration.
We assume that in a binary system, planetesimal formation follows
similar process as around a single star, and giant and terrestrial 
planets are formed through the interactions of these objects. It is 
important to emphasize that in such systems, the stellar companion has 
a strong effect on the accretion of planetesimals and the formation 
of larger bodies. In general, the perturbations due to the secondary star 
increase the relative velocities of planetesimals \citep{Hep78,Whitmire98}, 
which may cause their collisions to result in breakage and 
fragmentation. This object can also inhibit the formation of 
protoplanets by destabilizing  the regions where the building 
blocks of these objects exist \citep{Whitmire98}. It has, however, 
been shown that the effect of the binary companion on increasing 
the relative velocities of planetesimals may be counterbalanced
by dissipative forces such as gas-drag and dynamical friction
\citep{Marzari97,Marzari00,Thebault06}. As shown by these authors,
the combined effect of gas-drag and gravitational force of the
stellar companion results in a strong alignment of the periastra 
of planetesimals, which increases the efficiency of their accretion
by reducing their relative velocities. In this paper, we assume that 
a Jupiter-like planet has already formed around the primary of our 
binary star system\footnote{The formation of gas-giant planets in
a dual-star system is still a subject of research. While simulations 
by \citet{Nelson00} indicate that gas-giant planet formation may not 
proceed through the disk instability mechanism around the primary of 
a binary star system with separation of $\sim 50$AU, recent simulations 
by \citet{Boss06} show that Jupiter-like planets can form in such 
environments via the gravitational instability of a marginally 
unstable circumprimary disk. On the other hand, as shown by
\citet{Thebault04}, the core accretion mechanism may also
be able to form giant planets around the primary of a binary star system.
However, as the results of their simulations for planet formation in the 
$\gamma$ Cephei system indicate, the semimajor axis of the final gas-giant
planet may be smaller than its observed value.}, and the interactions
of planetesimals have been efficient and have resulted in the 
formation of a disk of planetary embryos (e.g., via oligarchic growth, 
cf. Kokubo \& Ida 1998) around this object.

As mentioned earlier, we focus our attention on the formation 
of habitable planets, that is, Earth-like objects in the habitable 
zone of the primary star. Since all life on Earth requires the 
presence of liquid water, we consider water-rich planets to be the 
best candidates for habitability, and pay close attention to the 
acquisition of water during the formation of these objects. 
Similar to the current models of the formation of habitable planets
in our solar system, we assume that cometary material, if existed 
around the stars of a binary system, would provide little to no 
water to the terrestrial planets that might form in the habitable 
zone of the primary star. We adopt the model of \citet{Morbidelli00}, 
who argued that water-rich bodies originating in the solar system's 
asteroid belt were the primary source of Earth's water, and simulate 
the late stage of terrestrial planet formation \citep{Wetherill96}
by numerically integrating the orbits of a few hundred protoplanetary 
objects, for different values of the mass, semimajor axis, and orbital 
eccentricity of the secondary star. We assume an initial gradient in the
water contents of protoplanets such that radial mixing is required to
"deliver" water to planets in the habitable zone \citep{Morbidelli00,
Raymond04,Raymond06a}. We identify the regions of the parameter-space 
of a binary-planetary system for which an Earth-like planet can form 
in the habitable region of the primary star.

The outline of this paper is as follows.
In $\S$ 2, we discuss the details of our model. Section 3 has to do with
the numerical integrations of the system and the analysis of the results. 
In $\S$ 4, we study the formation of habitable planets, and in $\S$ 5,
we conclude our study by reviewing the results and discussing 
their applications.

\section {The Model}

As mentioned in the introduction, we would like to study the formation of
terrestrial planets in the habitable zone of the primary of a 
moderately close binary-planetary system. We are primarily interested in
understanding how the motion of the secondary star affects the dynamics
of a disk of protoplanetary objects and the final assembly and 
water-contents of the resulted terrestrial-sized bodies. In other
words, we would like to study how the process of habitable planet
formation in a system consisting of a star, a disk of planetary embryos,
and a giant planet will be altered if a stellar companion is
introduced to the system. 

The statement above portrays a general picture of our model.
To ensure the habitability of such a system, Earth-like objects
have to form in the habitable zone of its primary star and
maintain long-term stable orbits in that region. On the other hand, 
as shown by \citet{Hagh06}, terrestrial planets can have stable orbits only 
at distances close to the primary star and outside the influence zone 
of its giant planet. This requires that 
the habitable zone of the primary to be considerably closer to 
it than the orbit of its planetary companion. To satisfy this requirement, 
and also for the purpose of comparing habitable
planet formation in binary-planetary systems with that around
single stars, we make the following assumptions.

\noindent
1) We assume that the primary of our system is a Sun-like star.
As indicated by \citet{Kasting93}, the habitable zone 
of such a star will extend from 0.95 AU to 1.37 AU. This is
a conservative estimate that places the outer boundary 
of the habitable zone at a distance where CO$_2$ clouds start 
to form \citep{Jones05}. The outer edge of this region
may in fact be at larger distances. As shown by \citet{Forget97}, 
and \citet{Mischna00}, the outer boundary of the habitable zone of the 
Sun may be at approximately 2.4 AU from this star. In this study,
we adopt a relatively moderate approach and consider a habitable 
zone between 0.9 AU and 1.5 AU for our primary star.

\noindent
2) We consider the giant planet of our system to be a 
1 Jupiter-mass object. Since we would like to study how the orbital 
dynamics of the secondary star affects the interactions of planetary 
embryos, we assume the orbit of this planet is circular. We also 
consider the semimajor axis of this object to be at 5 AU,
outside the habitable zone of the primary star.

\noindent
3) We choose the mass of the secondary star to be 0.5, 1.0, and 1.5 
solar-masses. We consider the semimajor axis of this object to have 
the values of 20, 30, and 40 AU, and its eccentricity to be 0, 0.2, 
and 0.4. Since we are interested in the formation of habitable 
planets at the presence of the Jovian-type planet of the system, 
it is necessary to ensure that, for any combination of these parameters,
the giant planet will have a long-term stable orbit . As shown by
\citet{Holman99}, in order for a planet in a circular orbit, to be 
stable in a binary star system, its semimajor axis cannot exceed the 
critical value $({a_c})$ given by
\begin{eqnarray}
&{{a_c}/{a_b}}=(0.464\pm 0.006)+
(-0.380 \pm 0.010){\mu_b} + (-0.631\pm0.034) {e_b}\nonumber\\
&\qquad\qquad\qquad
+ (0.586 \pm 0.061) {\mu_b} {e_b} + (0.150 \pm 0.041) {e_b^2} 
+(-0.198 \pm 0.074){\mu_b} {e_b^2}\>.
\end{eqnarray}
\vskip 2pt
\noindent
In this equation, $a_b$ and $e_b$
are the semimajor axis and orbital eccentricity of the
stellar companion, and ${\mu_b}={M_2}/{M_1}$, where
$M_1$ and $M_2$ are the masses of the primary and secondary stars,
respectively. Figure 1 shows the graph of $a_c$
in term of the eccentricity of the binary. In this figure, 
${\mu_b}=1$ and  ${a_b}=20,30$, and 40 AU.
As shown here,  a giant planet with a semimajor axis of 5 AU
will not have a stable orbit in an equal-mass binary
with a separation of 20 AU and an eccentricity of 0.2 or higher.
Similar situation exists for a 30 AU binary  with a 0.4 or larger
eccentricity. We use equation (1) to identify the combinations of
the mass and orbital parameters of the secondary star for which
the giant planet of the system becomes unstable, and simulate the formation
of habitable planets for those combinations of these parameters
that result in a stable orbit for this object.

\noindent
4) We assume that planetary embryos have already
formed in the region between the primary and the giant planet. 
We consider a disk of 115 Moon-to Mars-sized bodies,
with masses ranging from 0.01 to 0.1 Earth-masses,
randomly distributed, by 3 to 6 mutual Hill radii, between 0.5 AU and 4 AU 
from the primary star. The masses of embryos increase with their
semimajor axes $(a)$ and the number of their mutual Hill radii
$(\Delta)$ as ${a^{3/4}}{\Delta^{3/2}}$ \citep{Raymond04}.
The surface density of our disk model, normalized to 
a density of 8.2 g/cm$^2$ at 1 AU, varies as $r^{-1.5}$, where
$r$ is the radial distance from the primary star. The
total mass of our disk model is approximately 4 Earth-masses. 
Figure 2 shows the graph of one of such disks.

\noindent
5) We 
assume that the water to mass ratios  
of embryos follow the current distribution of water in
primitive asteroids of the asteroid belt \citep{Abe00}. That is,
embryos inside 2 AU are taken to be dry, the ones
between 2 to 2.5 AU are considered to contain 1\% water, and those
beyond 2.5 AU are assumed to have a water to mass ratio of 5\%
\citep{Raymond04,Raymond05a,Raymond05b,Raymond06a,Raymond06b}.
We also consider an initial Iron content for each embryo.
This value is obtained by interpolating between the values of the 
Iron contents of the terrestrial planets 
\citep{Lodders98,Raymond05a,Raymond05b}, with a dummy
value of 40\% in place of Mercury becuase of its anomalously high 
Iron content.

\section {Numerical Simulations}

Using the N-body integration package MERCURY \citep{Chambers99},
we numerically integrated the equations of motion of the planetary embryos
of our disk model for different values of the mass, semimajor axis, and
eccentricity of the secondary star. We allowed the protoplanetary
objects to collide with one another and assumed that each collision
was perfectly inelastic. We also assumed that no debris was generated 
during a collision, and that the effect of the energy released
during an impact, on the morphology and structure of 
the colliding objects, was negligible.

We carried out a total of 46 simulations, each with a time step 
of 6 days\footnote{Slightly smaller than 1/20 of the orbital period of the
closest embryo at 0.5 AU}. Figure 3 shows the results of one of 
such simulations. In this figure, the separation of the binary is 
30 AU, its eccentricity is 0.2, and the mass of the secondary star 
(not shown in the figure) is 0.5 solar-masses. As shown here, 
after 100 Myr, a terrestrial-sized object (1.1 Earth-masses), 
with substantial amount of water, is formed in the habitable 
zone of the primary star. The orbit of this object has
a semimajor axis of approximately 1.18 AU and an eccentricity of
$\sim 0.06$. The Jupiter-sized planet of the system is shown 
as a big black circle.

Similar to the simulations of terrestrial planet formation around 
single stars \citep{Morbidelli00,Raymond04}, our simulations are 
stochastic. That is, for a given set of orbital parameters of the 
binary companion, different initial distributions of protoplanetary 
objects produce different results. For this reason, for each set of the
initial orbital parameters of the binary companion, we carried out 
simulations for three different random distributions of planetary embryos.
Figures 4 shows the final result of such simulations for two different 
cases. The case on the left corresponds to a binary with a separation 
of 30 AU and a secondary of 1.0 solar-mass in a circular orbit.
The case on the right represents the results of simulations in a 
system in which the same secondary star is now in an orbit with a 
semimajor axis of 40 AU and an eccentricity of 0.2. Each simulation, 
from top to bottom, corresponds to different distribution of planetary 
embryos with the simulations on the same row having similar initial 
distributions of protoplanetary objects.

In addition to the stochasticity of simulations, figure 4 also shows the 
relation between the orbital eccentricity of the stellar companion and the
water contents of the final bodies. As shown here, 
for identical initial distributions of planetary embryos
(i.e., simulations on the same rows), the total water content of 
the system on the left, where the secondary star is in a circular
orbit, is higher than that of the system on the right, where the 
orbit of the secondary is eccentric. Figure 5 shows this for several 
other simulations. As depicted by this figure, for a given 
separation of the binary, the accumulative water content of the final planets
decreases as the eccentricity of the binary becomes larger.

The fact that the final assembly of terrestrial planets in a system with 
an eccentric secondary star contains less water implies that, prior to 
the formation of these objects, most of the water-carrying embryos have 
left the system. Our simulations indicate that on average 90\% of 
embryos in these systems were ejected during the course of the integration 
(i.e., either their semimajor axes exceeded 100 AU, or their orbital 
eccentricities became larger than unity) and among them, 60\% 
collided with other protoplanetary bodies prior to their ejection 
from the system. A small fraction of embryos $(\sim 5\%)$ also 
collided with the primary or secondary star, or with the Jupiter-like 
planet of the system.

The destabilizing effect of an eccentric secondary star in a binary
system has also been reported by \citet{Artymowicz94}, and
\citet{David03}. As shown by these authors, 
in binaries with small perihelia, the interactions of small bodies
with the secondary star shorten the lifetimes of these objects 
and enhance the disk truncation. In binary-planetary systems, 
in addition to the perturbation from the stellar companion,
similar to our solar system, 
\citep{Chambers02-II,Levison03,Raymond04,Raymond06a}, 
planetesimals and protoplanetary objects are also subject to
the perturbative effect of the giant planet of the system.
In such systems, the Jovian-type planet plays the important role of 
transferring angular momentum from the secondary star to planetary 
embryos and strongly affects the motion and radial mixing of these objects. 
Figure 6 shows this in more details. The systems simulated here are 
binaries with  0.5 solar-masses secondary stars and separations 
of 30 AU. The binary eccentricity in these systems is equal 
to 0, 0.2 and 0.4, from top to bottom. As shown here, as the 
eccentricity  of the binary companion increases,
its perihelion becomes smaller and its interaction with the
giant planet becomes stronger. The latter
causes the eccentricity of the giant body to increase and 
results in its closer approach to the disk of planetary embryos and
enhancing collisions and mixing among these objects. The 
eccentricities of embryos, at distances close to the outer edge of the 
protoplanetary disk, rise to higher values until these bodies are
ejected from the system. In binaries with smaller perihelia, 
the process of the transfer of angular momentum by means of the 
giant planet is stronger and the ejection of protoplanets occurs 
at earlier times.
As a result, the total water to mass ratios of such systems
become smaller as the eccentricities of their stellar companions increase.

To further study the effect of the stellar companion on the
dynamics and radial mixing of embryos, we carried out several 
simulations without the Jupiter-like planet of the system.
Results indicate that, it is indeed possible to form terrestrial-class
planets, with significant amounts of water, in the habitable
zone of the primary star. However, because of the lack of the
intermediate effect of the Jovian-type planet, the interaction of
embryos is slower and terrestrial planet formation 
takes longer. Figure 7 shows this for three systems.
The separation of the binary in each system is 30 AU,
and the mass of the secondary star is 1 solar-mass. 
The eccentricity of the secondary star is equal to 0, 0.2, and
0.4 in simulations from left to right, respectively.

An interesting result depicted by figure 7 is the decrease in
the number of the final terrestrial planets, and increase in their sizes and 
accumulative water contents with increasing the eccentricity of 
the secondary star. The simulation on the left, in which the 
secondary is in a circular orbit, shows that, since in this system, 
the transfer of angular momentum from the stellar companion to 
protoplanetary objects, by means of the Jupiter-like planet of the system, 
is non-existent, the radial mixing of embryos is
slow and inefficient. In binaries with larger eccentricities,
the close approach of the stellar companion to the disk of
protoplanets increases the rate of the interaction of these objects 
and enhances their collisions and radial mixing.
As a result, in such systems, more of the water-carrying embryos
participate in the formation of the final terrestrial planets. 
It is important to emphasize that, as explained below, 
this process is efficient only in moderately eccentric binaries.
In binary systems with high eccentricities (small perihelia), 
embryos may be ejected from the system \citep{David03}, and 
terrestrial planet formation may become inefficient.

An important result shown by figure 7 is the existence of
a trend between the binary perihelion distance and the location 
of the outermost terrestrial planet. Figure 8 shows this for a 
set of different simulations. The top panel in this
figure represents the semimajor axis of the outermost 
terrestrial planet, $a_{out}$, as a function of the binary 
eccentricity, $e_b$. The bottom panel shows the ratio of 
this quantity to the perihelion distance of the binary, $q_b$.
As shown here, simulations with no giant planet (shown in black) 
follow a clear trend: Terrestrial planets only form interior to 
roughly 0.19 times the binary perihelion distance. This has also 
been noted by \citet[][see their figure 9]{Quintana07} in their simulations
of terrestrial planet formation in close binary star systems. 
The smallest binary perihelion that allows terrestrial planets to
form outside the inner edge of the habitable zone (0.9 AU) in these 
systems is simply 0.9/0.19 = 4.7 AU, comparable to the estimate 
by \citet{Quintana07}. In binaries with no giant planets, 
Sun-like primaries with companions with perihelion distances 
smaller than approximately 5 AU are therefore not good candidates 
for habitable planet formation. It is, however, important to note that, 
because the stellar luminosity, and therefore the location of the 
habitable zone, are sensitive to stellar mass 
\citep{Kasting93,Raymond07b}, the minimum binary
separation necessary to ensure habitable planet formation 
will vary significantly with the mass of the primary star.

In simulations with giant planets, figure 8 indicates  that 
terrestrial planets form closer-in. The ratio $a_{out}/q_b$ in these 
systems varies between approximately 0.06 and 0.13, depending on
the orbital separation of the two stars. The accretion process in such systems 
is more complicated since the giant planet's eccentricity and 
its ability to transfer angular momentum are largely regulated 
by the binary companion.    

\section{Habitable Planet Formation}

Despite the stochasticity of the simulations and the
large size of the parameter-space, many of our integrations 
resulted in the formation of Earth-sized objects, 
with substantial amounts of water,
in the habitable zone of the primary star.
Figures 9 and 10 show the results of some of these simulations. 
The orbital parameters of the final objects and their water contents
are listed in Table 1. It is important to mention that,
in comparing the water contents of the Earth-like planets
of our simulations with those of the Earth, we consider
the water to mass ratio of the Earth to be 0.001. Since the exact amount
of water in the Earth's mantle is unknown (between 1-10 Earth's
ocean), such an estimate of Earth's water-mass fraction is 
equivalent to considering one ocean of water 
$(\sim 1.5 \times 10^{24}\, {\rm g})$ on the Earth's surface 
and three oceans in its mantle.

A detailed analysis of the results depicted by figures 9 and 10
indicate that the systems shown in these figures have relatively
large perihelia. Figure 11 shows this for simulations with
${\mu_b}=1$, in terms 
of the semimajor axis and eccentricity of the stellar companion.
The circles in this figure represent those systems whose parameters
were chosen from figure 1 (i.e., stable giant planets) and their 
simulations resulted in the formation of habitable bodies. 
The number associated with each circle corresponds to
the mean eccentricity of the giant planet during the simulation. 
For the sake of comparison, the systems in which
the giant planet is unstable have also been marked.
Given that at the beginning of each simulation,
the orbit of this planet was considered to be
circular, a non-zero value for its average eccentricity 
is the result of its interaction with the secondary star.
The fact that Earth-like objects were formed in systems
where the average eccentricity of the giant planet is small implies that
this interaction has been weak. In other words,
binaries with moderate to large perihelia and with giant
planets on low eccentricity orbits are most favorable
for habitable planet formation.
Similar to the formation of habitable planets around single stars,
where giant planets, in general, play destructive roles, 
a strong interaction
between the secondary star and the giant planet in a binary-planetary system 
(i.e., a small binary perihelion) increases the orbital
eccentricity of this object, and results in the removal of the terrestrial 
planet-forming materials from the system.

\section {Summary and Conclusions}

We presented the results of the numerical simulations of the formation
of Earth-like bodies in the habitable zones of moderately close
binary-planetary system. The systems of our interest had binary
separations equal to or smaller than 40 AU, their primary stars hosted
Jupiter-like planets, and their habitable zones were closer
to their primaries than the orbits of their Jupiter-like objects.
We simulated the late stage of terrestrial planet formation 
in such systems, and studied the effect of the binary
companion on the interaction of planetary embryos and
the formation, dynamical evolution, and
physical characteristics of the final terrestrial bodies.
Our results indicate that, terrestrial-class planets
can form in the habitable zones of binary-planetary 
systems, and their final assembly, as well as their
water contents, strongly depend on the separation of the binary
and the eccentricity of the stellar companion.

Our simulations show that, similar to the formation of
Earth-like objects at the presence of Jupiter in our solar system, where
the eccentricity of the giant planet becomes the key 
factor in delivering water to the habitable zone 
\citep{Chambers02-II,Raymond04}, the orbital dynamics of
this object, and its intermediate role in
transferring the perturbative effect of the secondary 
star to protoplanetary objects, in a binary planetary system,
have strong effects on the radial mixing of embryos 
and the water-delivery process. As shown by \citet{Raymond07a},  
giant planets are, in general, unfavorable for delivery of water 
from asteroidal regions to the habitable zone. In a binary system,
the interaction between this object and the secondary 
star, particularly in eccentric binaries, increases its orbital 
eccentricity and makes the process of water delivery inefficient. 
Our calculations show that, to form habitable planets at the presence 
of a Jupiter-like planet in a binary star system, this interaction 
has to be weak. Water delivery is more efficient when the perihelion 
of the binary is large \citep{Quintana07} and the orbit of the 
giant planets is close to a circle \citep{Raymond06a}.

As mentioned in the introduction, this study was based on 
two fundamental assumptions: (1) terrestrial planet formation in 
binary-planetary systems follows the same process as in single stars, 
(2) planetary embryos have already formed through oligarchic growth, and
terrestrial-class objects are formed through the gravitational interaction
of these bodies. Despite the observations of 
circumstellar disks around the components of several binary star systems, 
such assumptions may not be entirely viable.
It is, in fact, necessary to study how the presence of a stellar companion
affects the formation of planetary embryos, and 
also how the chemical structure of a disk of protoplanetary objects
changes at the presence of a second close-by star. 

The final water contents of the terrestrial planets is also an issue
that requires detailed considerations. In our simulations, we assumed
that all collisions were perfectly inelastic, and the water contents
of the resulted planet would be equal to the sum of the water contents
of the impacting bodies. This is an assumption that sets an upper limit
for the water budget of terrestrial planets, and ignores the loss
of water due to the impact. In fact, as shown by  \citet{Genda05}
and \citet{Canup06}, portions of the water of the impacting bodies
will be lost due to the impact and the motion of the ground of 
an impacted body. The total water budget of the final bodies
of figures 9 and 10 may in fact be 5-10 times smaller than those 
reported in Table 1 \citep{Raymond04}.
  
The simulations presented here are low resolution. Our model
includes only large objects. A more realistic model 
would contain smaller bodies such as km-sized planetesimals, as well.
For this reason, the results of our integrations may not reveal
detail characteristics of the final planetary systems, as they would
be obtained from high resolution simulations 
[e.g., see \citet{Raymond06c}]. However, as shown
by \citet{Agnor99} and \citet{Chambers01} for the simulations of our
solar system, such integrations can produce the main general properties
of the final assembly of the planetary bodies. That, combined with the fact 
that the speed of computational simulations varies as $N^2$,
where $N$ is the number of involved objects, makes the
low resolution simulations ideal for exploring the system's 
parameter-space.

Despite of its limitations, our study shows that habitable
planet-formation can be efficient in moderately close
binary-planetary systems. The favorable systems seem to be binaries 
with moderate to large perihelia and giant planets outside
the habitable zones of their primary stars and in low eccentricity 
orbits. This points the fact that, to develop a more comprehensive 
understanding of the formation of habitable planets in such environments,
it is important to obtain a better knowledge of the formation of 
giant planets in these systems.

\acknowledgments

We are thankful to the Department of Terrestrial Magnetism at the
Carnegie Institution of Washington for access
to their computational facilities where the numerical simulations
of this work were performed.
This work has been supported by the NASA Astrobiology
Institute under Cooperative Agreement NNA04CC08A at the Institute 
for Astronomy at the University of Hawaii-Manoa for NH. SNR was
supported by an appointment to the NASA Postdoctoral Program
and the University of Colorado Astrobiology Center, administered by Oak
Ridge Associated Universities through a contract with NASA.

\clearpage

\begin{figure}
\plotone{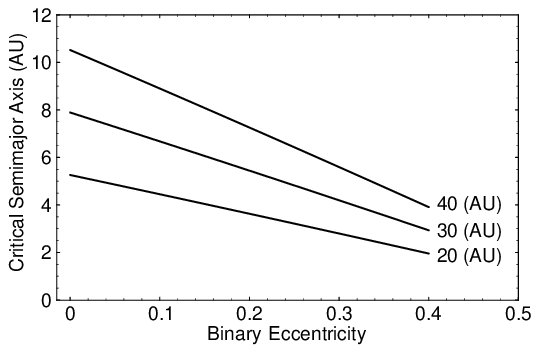}
\vskip -3.3in
\caption{Graph of the critical semimajor axis of a planet in
an equal-mass binary star system for three different values 
of the separation of the binary.
\label{fig1}}
\end{figure}

\clearpage

\begin{figure}
\plotone{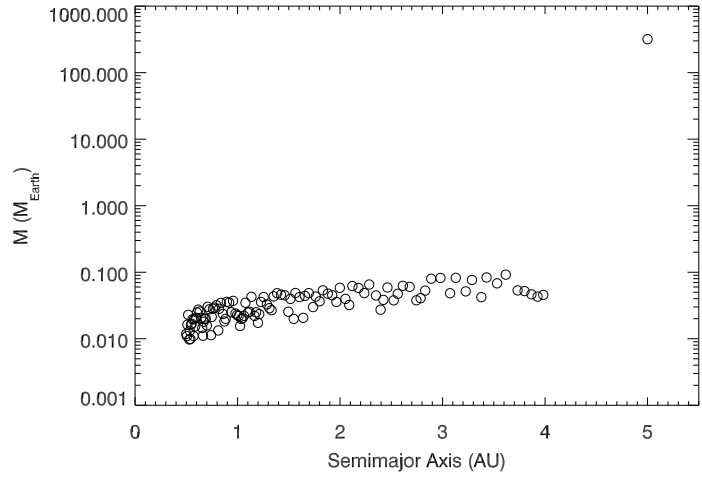}
\vskip -2in
\caption{Radial distribution of original protoplanetary objects.
\label{fig2}}
\end{figure}

\clearpage

\begin{figure}
\plotone{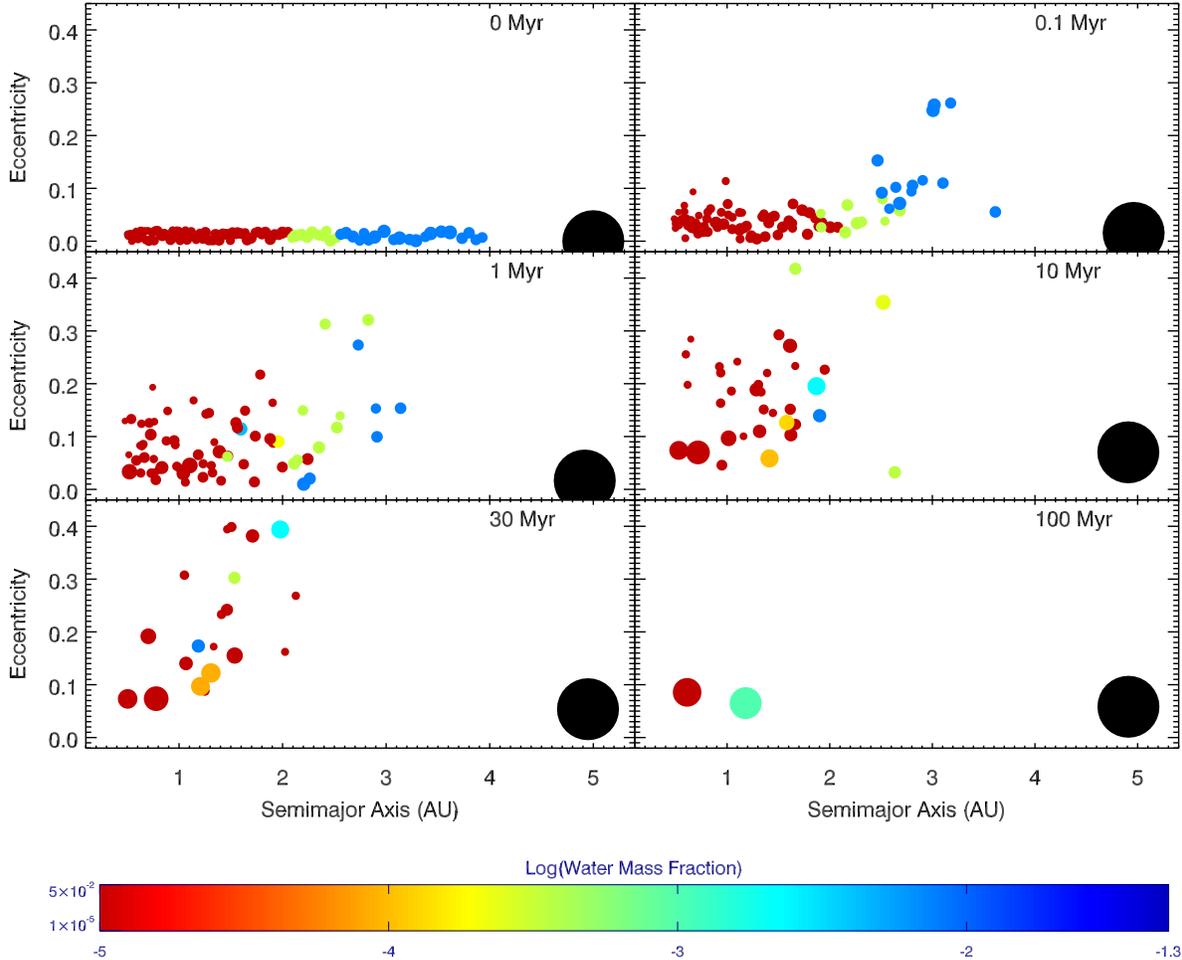}
\vskip 20pt
\caption{Snap shots of the
interaction of protoplanetary objects and the formation
of terrestrial planets. The mass of the secondary star is 0.5
solar-masses and its semimajor axis and eccentricity are 30 AU
and 0.2, respectively. The results show a terrestrial-sized planet,
with substantial amounts of water, at a semimajor axis of 1.2 AU and with an 
eccentricity of approximately 0.07. The Jupiter-sized planet
of the system is shown by the black circle.
\label{fig3}}
\end{figure}

\clearpage
\begin{figure}
\plotone{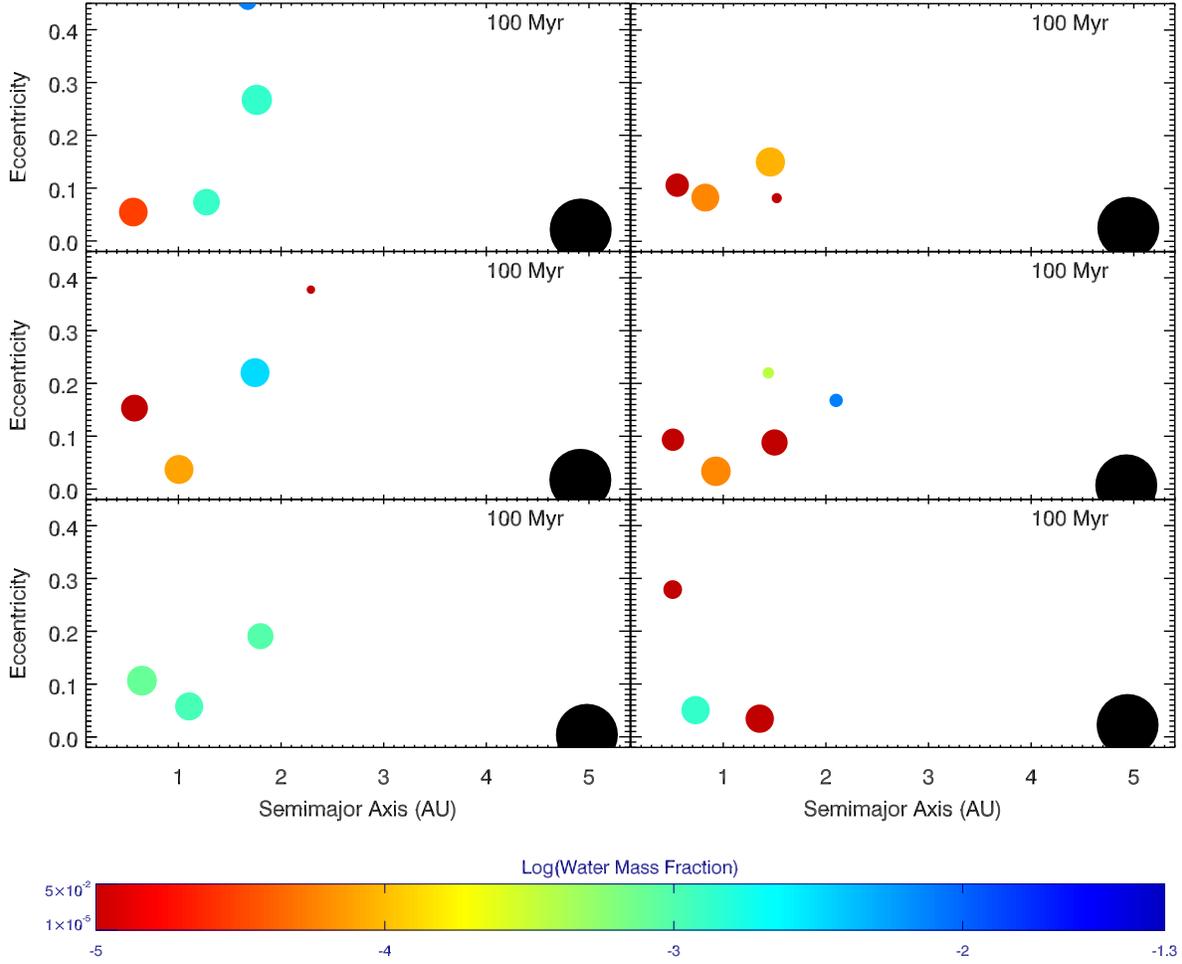}
\vskip 30pt
\caption{Stochasticity of simulations. The left column depicts the
results of three simulations with random distribution of protoplanetary
objects in a binary system with a Sun-like star as its secondary and 
in a circular orbit with a radius of 30 AU. The right column shows 
the results of
simulations for similar distributions of planetary embryos in a system
with a 1.0 solar-mass secondary star in an orbit with a semimajor
axis of 40 AU and eccentricity of 0.2. As shown in each column, different
distribution of planetary embryos may result in the formation of
different number of planets with different water to mass ratios.
\label{fig4}}
\end{figure}

\clearpage
\begin{figure}
\plotone{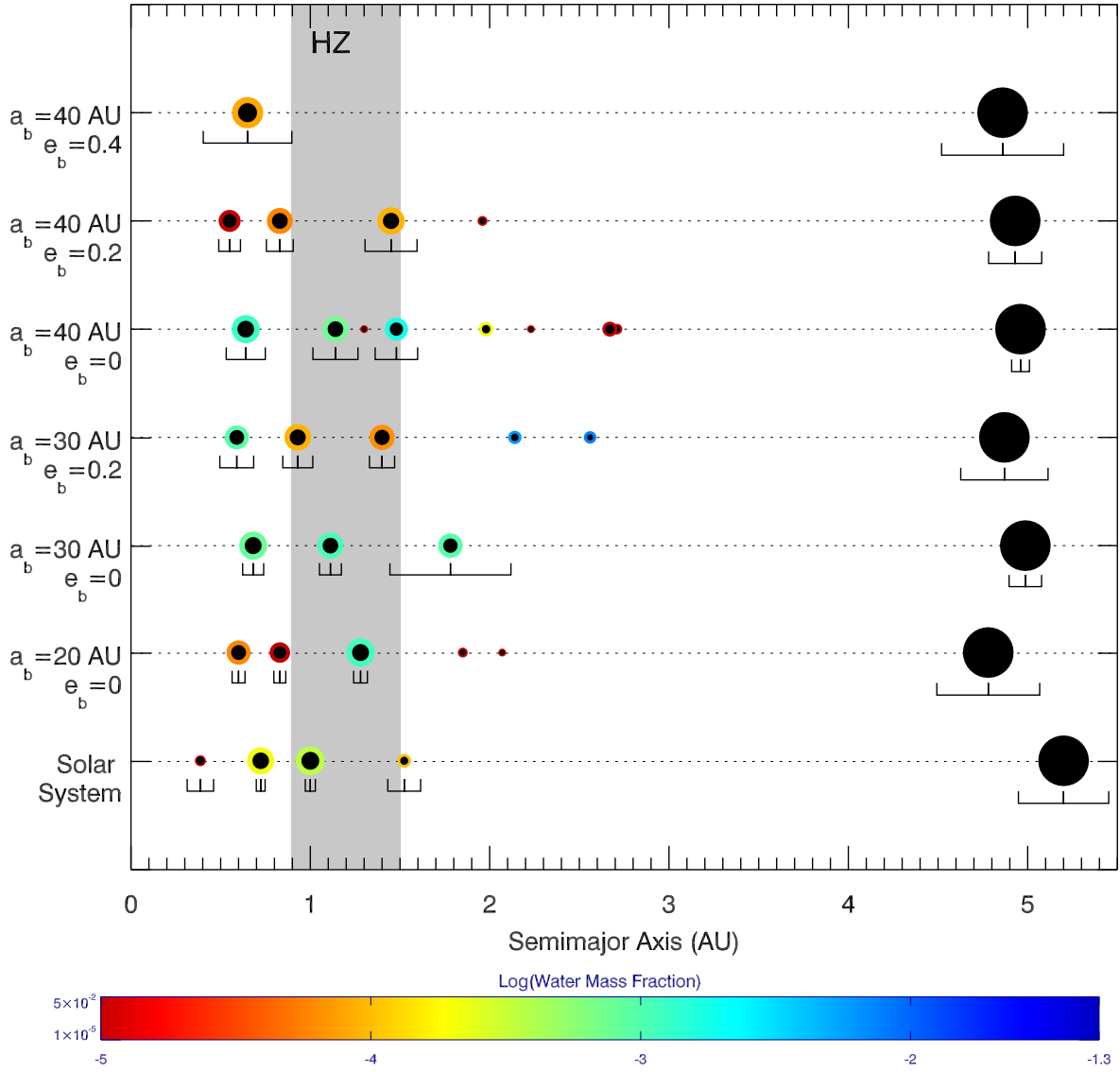}
\caption{Results of simulations in a system with ${\mu_b}=1$
for different values of the
eccentricity $(e_b)$ and semimajor axis $(a_b)$ of the stellar companion. 
\label{fig5}}
\end{figure}

\clearpage
\begin{figure}
\plotone{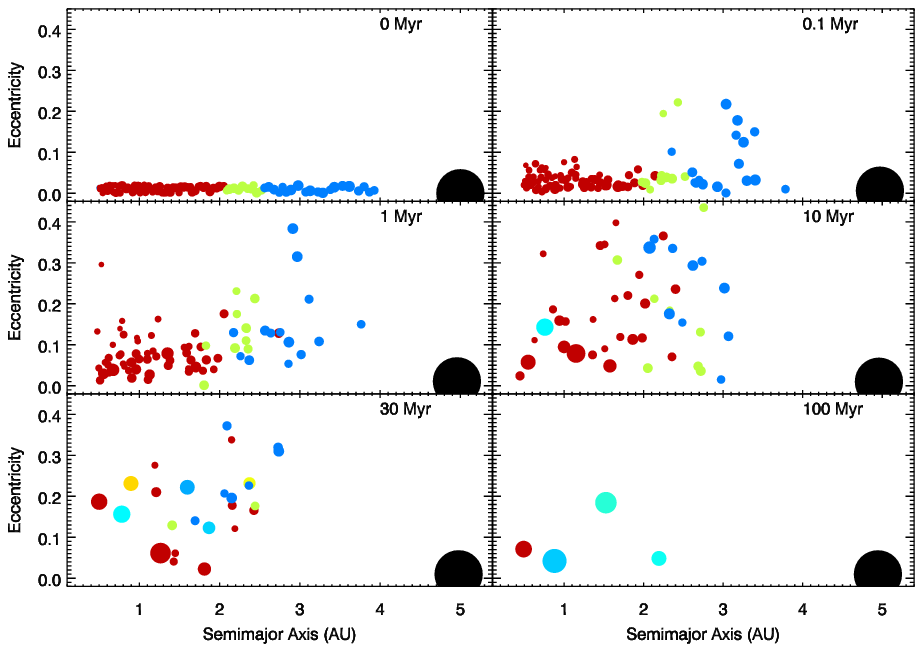}
\plotone{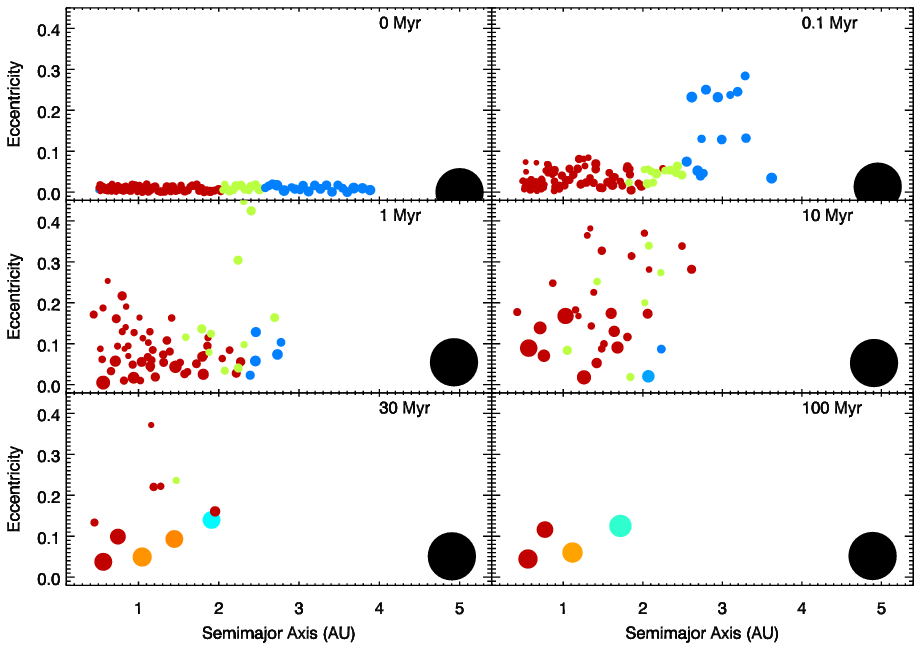}
\plotone{f6c.eps}
\vskip -6in
\caption{Variation of water contents with the eccentricity 
of the stellar companion. The primary star has a mass of 0.5
solar-masses, the semimajor axis of the binary is 30 AU,
and its eccentricity is equal to 0,0.2, and 0.4, from top to bottom.
\label{fig6}}
\end{figure}

\clearpage
\begin{figure}
\plotone{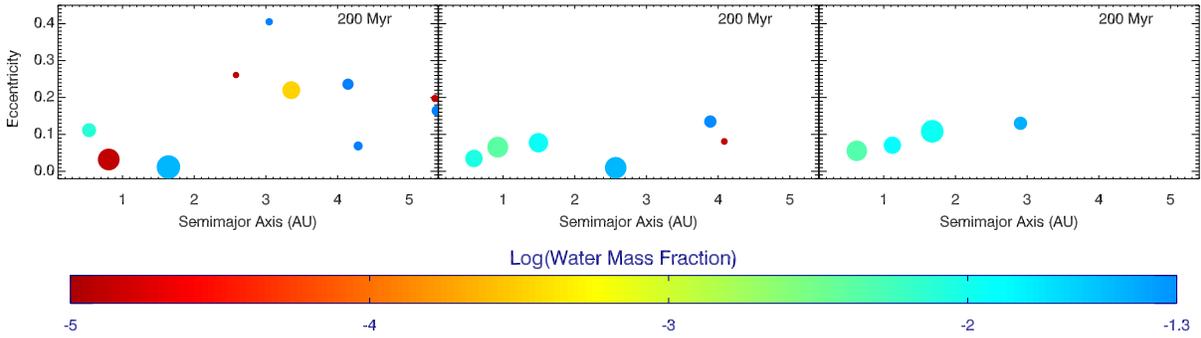}
\vskip -5.5in
\caption{Results of simulations in binary systems with no 
Jupiter-like planet. The stars of each binary are Sun-like
and their separations are 30 AU. The orbital eccentricity
of the secondary star is 0, 0.2, and 0.4, for the systems on 
the left, middle, and right, respectively. Note that the time
of integration has been increase to 200 Myr. 
\label{fig7}}
\end{figure}

\clearpage
\begin{figure}
\plotone{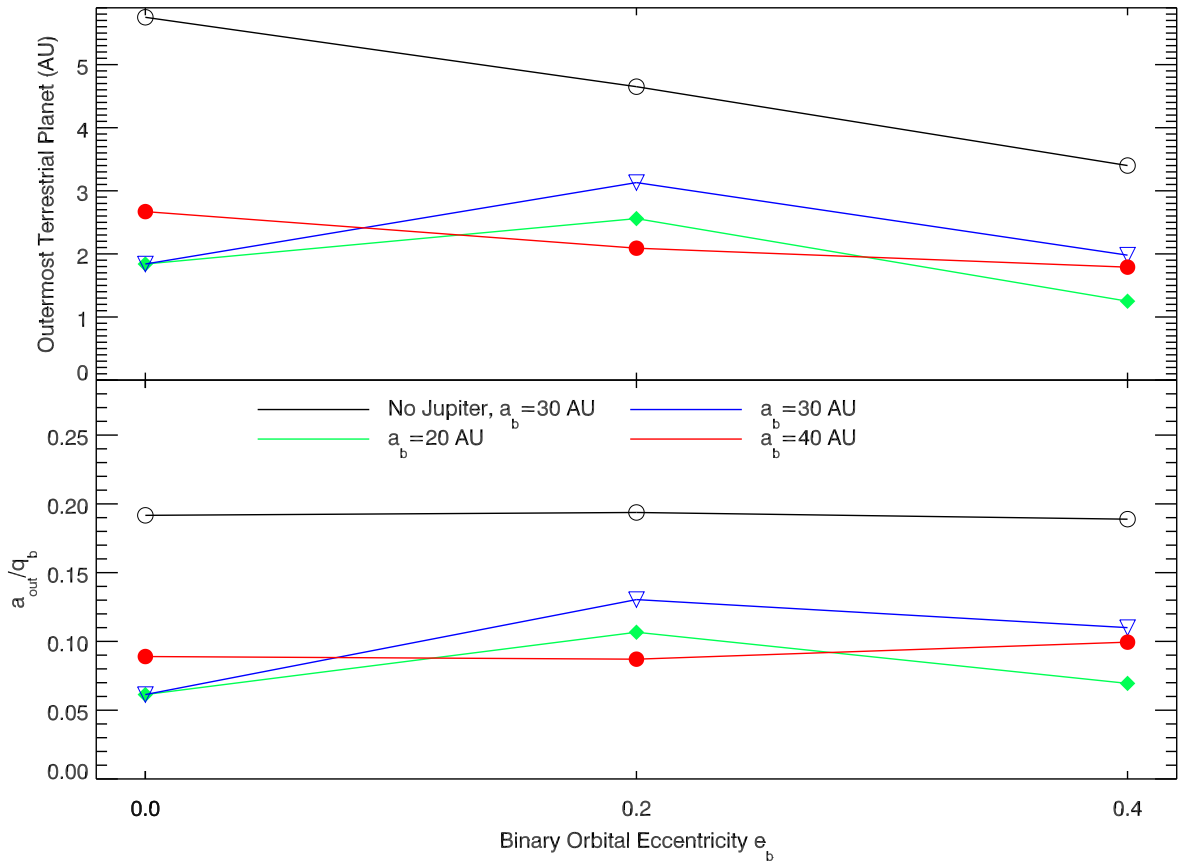}
\caption{Top panel: Semimajor axis of the outermost terrestrial planet.
Bottom panel: The ratio of the semimajor axis of this object 
to the perihelion distance of the binary. The secondary star is solar mass.
\label{fig8}}
\end{figure}

\clearpage
\begin{figure}
\plotone{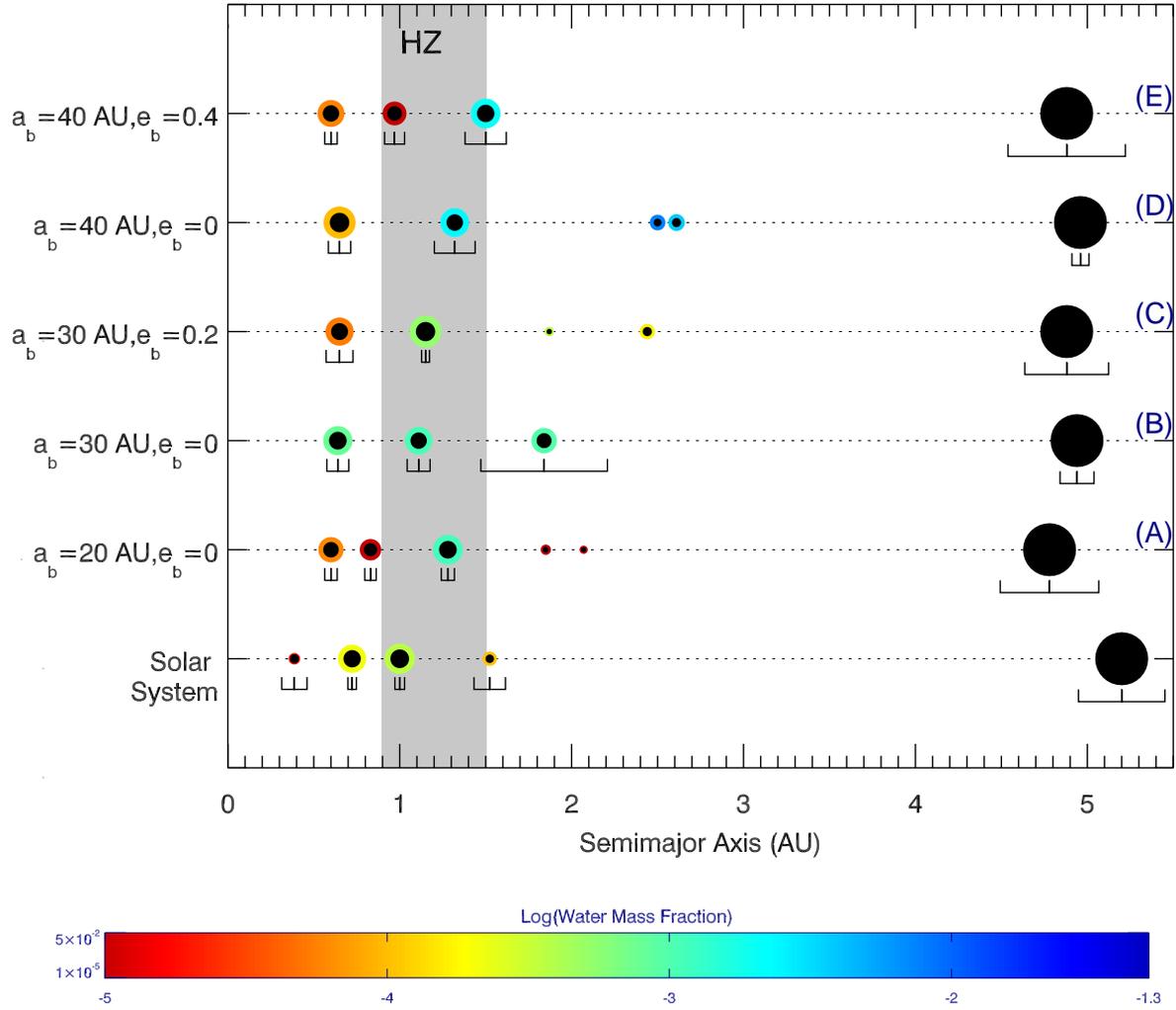}
\caption{Earth-like objects in the habitable zone of the primary
star. The mass of the secondary star in all simulations is 1 solar-mass.
\label{fig9}}
\end{figure}

\clearpage
\begin{figure}
\plotone{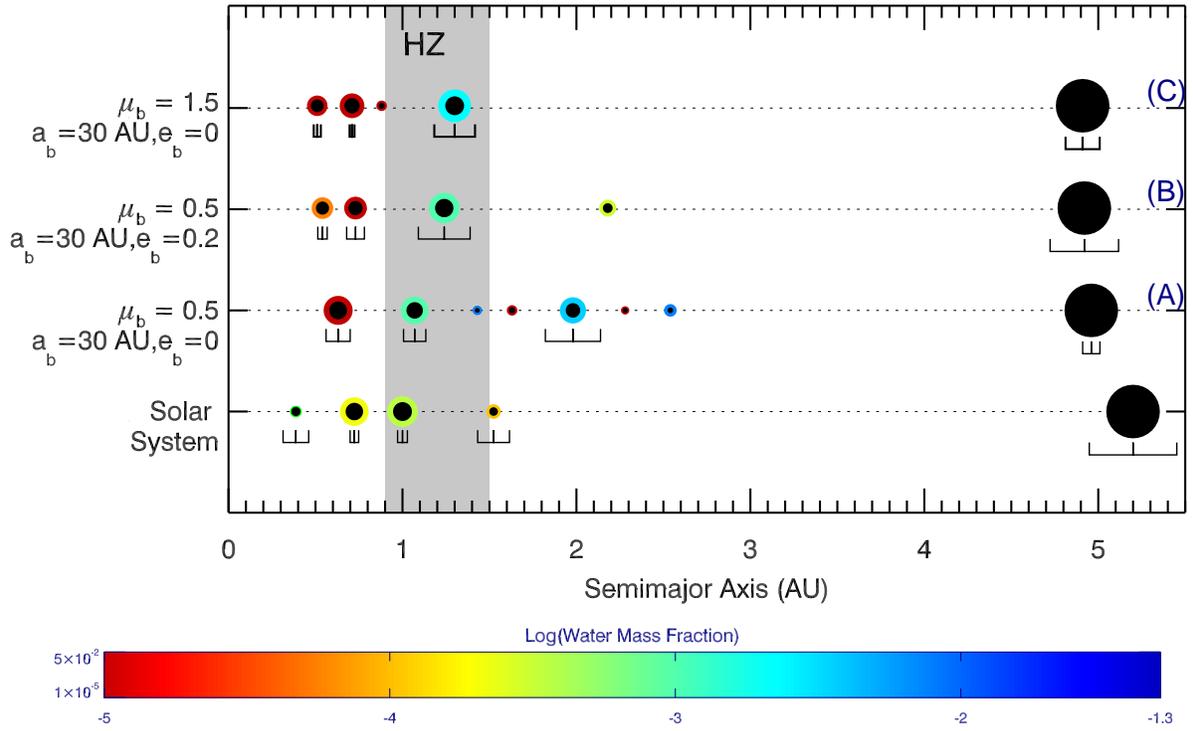}
\caption{Habitable planet formation in binary-planetary systems
with 0.5 and 1.5 solar-masses secondary stars.
\label{fig10}}
\end{figure}

\clearpage
\begin{figure}
\plotone{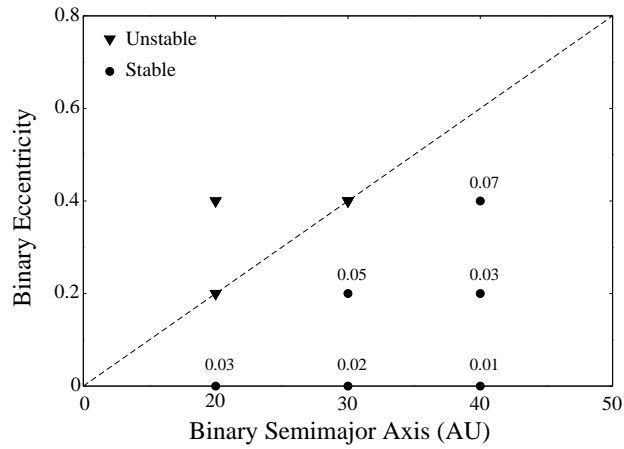}
\vskip -2.5in
\caption{Habitable planet formation in the $({e_b},{a_b})$ space
of an equal-mass binary-planetary system.
Circles correspond to binaries with initial parameters chosen
from figure 1, in which habitable planets are formed. Triangles
represent systems in which the giant planet is unstable.
The number associated with each circle represents the mean
eccentricity of the giant planet of the system during the simulation.
\label{fig11}}
\end{figure}

\clearpage
\begin{deluxetable}{lcccc} 
\tablewidth{0pt} 
\tablecaption{Characteristics of the Earth-like planets of figures 9 and 10.
\hfill
\vskip 1pt
(Sun's habitable zone: 0.9 - 1.50 AU)}
\tablehead{ 
\colhead{Simulation} &
\colhead{Planet Mass $(M_\oplus)$}  &
\colhead{Semimajor Axis (AU)} &
\colhead{Eccentricity} &
\colhead{Water/Mass}}
\startdata 
8-A   & 0.95    & 1.28     & 0.03    & 0.00421     \\
8-B   & 0.75    & 1.11     & 0.06    & 0.00415     \\
8-C   & 1.17    & 1.16     & 0.03    & 0.00164      \\
8-D   & 0.86    & 1.33     & 0.09    & 0.01070       \\    
8-E   & 0.95    & 1.50     & 0.08    & 0.00868       \\
9-A   & 0.74    & 1.07     & 0.06    & 0.00349       \\
9-B   & 0.99    & 1.26     & 0.12    & 0.00366       \\
9-C   & 1.23    &1.30      & 0.09    & 0.00103       \\

\enddata 
\end{deluxetable}


\begin{thebibliography}{}
\bibitem[Abe et al. (2000)]{Abe00}
Abe, Y., Ohtani, E., Okuchi, T., Righter, K., \& Drake, M. 2000,
in Origin of the Earth and the Moon, ed. K. Righter \& R. Canup
(Tucson: Univ. Arizona Press), 413
\bibitem[Agnor et al. (1999)]{Agnor99}
Agnor, C. B., Canup, R. M., \& Levison, H. F. 1999, Icarus, 142, 219
\bibitem[Akeson et al. (1998)]{Akeson98}
Akeson, R. L., Koerner, D. W., Jensen, E. L. N. 1998, \apj, 505, 358
\bibitem[Artymowicz \& Lubow (1994)]{Artymowicz94}
Artymowicz, P., \& Lubow, S. H. 1994, \apj, 421, 651 
\bibitem[Barbieri et al. (2002)]{Barbieri02}
Barbieri, M., Marzari, F., $\&$ Scholl, H. 2002, A\&A 396, 219
\bibitem[Boss (2006)]{Boss06}
Boss, A. P. 2006, \apj, 641, 1148
\bibitem[Canup \& Pierazzo (2006)]{Canup06}
Canup, R. M., \& Pierazzo, E. 2006, 
37th Annual Lunar and Planetary Science Conference, 2146
\bibitem[Chambers (1999)]{Chambers99}
Chambers, J. E. 1999, MNRAS, 304, 793 
\bibitem[Chambers (2001)]{Chambers01}
Chambers, J. E. 2001, Icarus, 152, 205
\bibitem[Chambers \& Cassen (2002)]{Chambers02-II}
Chambers, J. E., \& Cassen, P. 2002, M\&PS, 37, 1523
\bibitem[David et al. (2003)]{David03}
David, E., Quintana, E. V., Fatuzzo, M., \&  Adams, F. C., 
2003, \pasp, 115, 825 
\bibitem[Els et al. (2001)]{Els01}
Els, S. G., Sterzik, M. F., Marchis, F., Pantin, E., Endl, M.,
$\&$ K\"urster, M. 2001, A$\&$A, 370, L1
\bibitem[Forget \& Pierrehumbert (1997)]{Forget97}
Forget, F., \& Pierrehumbert, R. T. 1997, Science, 278, 1273
\bibitem[Genda \& Abe (2005)]{Genda05}
Genda, H., \& Abe, Y. 2005, 
36th Annual Lunar and Planetary Science Conference, 2265
\bibitem [Haghighipour (2006)]{Hagh06}
Haghighipour, N. 2006, \apj, 644, 543
\bibitem[Hatzes et al.(2003)]{Hatzes03}
Hatzes, A. P., Cochran, W. D., Endl, M., McArthur, B.,
Paulson, D. B., Walker, G. A. H., Campbell, B., $\&$ Yang, S.
2003, \apj, 599, 1383
\bibitem[Hayashi (1981)]{Hayashi81}
Hayashi, C. 1981, Prog. Theor. Phys. Suppl., 70, 35
\bibitem[Heppenheimer (1978)]{Hep78}
Heppenheimer, T. A. 1978, A\&A, 65, 421
\bibitem[Holman \& Wiegert (1999)]{Holman99}
Holman, M. J., \& Wiegert, P. A. 1999, \aj, 117, 621
\bibitem[Jones, Underwood, $\&$ Sleep (2005)]{Jones05}
Jones, B. W., Underwood, D. R., $\&$ Sleep, P. N. 2005, \apj, 622, 1091
\bibitem[Kokubo \& Ida (1998)]{Ida98}
Kokubo, E., \& Ida, S. 1998, Icarus, 131, 171
\bibitem[Kasting, Whitmire \& Reynolds (1993)]{Kasting93}
Kasting, J. F., Whitmire, D. P., $\&$ Reynolds, R. T. 1993,
Icarus, 101,108
\bibitem[Konacki (2005)]{Konacki05}
Konacki, M. 2005, Nature, 436, 230
\bibitem[Levison \& Agnor (2003)]{Levison03}
Levison, H. F., \& Agnor, C. 2003, \aj, 125, 2692
\bibitem[Lissauer et al. (2004)]{Lissauer04}
Lissauer, J. J., Quintana, E. V., Chambers, J. E., Duncan, M. J., $\&$
Adams, F. C. 2004, RevMexAA (Series de Conferencias), 22, 99
\bibitem[Lodders \& Fegley (1998)]{Lodders98}
Lodders, K., \& Fegley, B., Jr. 1998, 
Meteoritics \& Planetary Science, 33, 871
\bibitem[Marzari et al. (1997)]{Marzari97}
Marzari, F., School, H., Tomasella, L., \& Vanzani, V. 1997,
Planet.Space.Sci., 45, 337
\bibitem[Marzari $\&$ Scholl (2000)]{Marzari00}
Marzari, F., $\&$ Scholl, H. 2000, \apj, 543, 328
\bibitem[Mathieu (1994)]{Math94}
Mathieu, R. D. 1994, Annu. Rev. Astron. Astrophys., 32, 465
\bibitem[Mathieu et al. (2000)]{Math00}
Mathieu, R. D., Ghez, A. M., Jensen, E. L. N., $\&$
Simon, M. 2000, in Protostars and Planets IV, ed. V. Mannings,
A. P. Boss, $\&$ S. S. Russell (Tucson: Univ. Arizona Press), 703
\bibitem[Mischna et al. (2000)]{Mischna00}
Mischna, M. A., Kasting, J. F., Pavlov, A., \& Freedman, R.
2000, Icarus, 145, 546
\bibitem[Morbidelli et al. (2000)]{Morbidelli00}
Morbidelli, A., Chambers, J., Lunine, J, I.,  Petit, J. M.,
Robert, F., Valsecchi, G., B., Cyr, K., E. 2000, 
Meteorit. Planet. Sci., 35, 1309
\bibitem[Nelson (2000)]{Nelson00}
Nelson, A. F. 2000, \apj, 537, L65
\bibitem[Quintana et al. (2002)]{Quintana02}
Quintana, E. V., Lissauer, J. J., Chambers, J. E., $\&$ Duncan, M. J.
2002, \apj, 576, 982
\bibitem[Quintana (2003)]{Quintana03}
Quintana, E. V., 2003, in Scientific Frontiers in Search for
Extrasolar Planets, eds. D. Deming, \& Seager, S., ASP Conference Series,
294, 319 
\bibitem[Quintana \& Lissauer (2006)]{Quintana06}
Quintana, E. V., \& Lissauer, J. J., 2006, Icarus, 185, 1
\bibitem[Quintana et al. (2007)]{Quintana07}
Quintana, E. V., Adams, F. C., Lissauer, J. J., \& Chambers, J. E.,
2007, to appear in \apj (astro-ph/0701266)
\bibitem[Raymond et al. (2004)]{Raymond04}
Raymond, S. N., Quinn, T., \& Lunine, J., I. 2004, Icarus, 168, 1
\bibitem[Raymond et al. (2005a)]{Raymond05a}
Raymond, S. N., Quinn, T., \& Lunine, J., I. 2005a, \apj, 632, 670
\bibitem[Raymond et al. (2005b)]{Raymond05b}
Raymond, S. N., Quinn, T., \& Lunine, J., I. 2005b, Icarus, 177, 256
\bibitem[Raymond (2006)]{Raymond06a}
Raymond, S. N. 2006, \apjl, 643, L131
\bibitem[Raymond, Barnes \& Kaib (2006)]{Raymond06b}
Raymond, S. N., Barnes, R., \& Kaib, N. A. 2006, \apj, 644, 1223
\bibitem[Raymond, Mandell \& Sigurdsson (2006)]{Raymond06c}
Raymond, S. N., Mandell, A. M., \& Sigurdsson, S. 2006, Science, 313, 1413 
\bibitem[Raymond (2007)]{Raymond07a}
Raymond, S. N. 2007, in preparation
\bibitem[Raymond et al. (2007)]{Raymond07b}
Raymond, S. N., Quinn, T., \& Lunine, J.I. 2007,  Astrobiology, in press,
(astro-ph/0510285)
\bibitem[Rodriguez et al. (1998)]{Rodriguez98}
Rodriguez, L. F., D'Alessio, P., Wilner, D. J., Ho, P. T. P., 
Torrelles, J. M., Curiel, S., Gomez, Y., Lizano, S., Pedlar, A., 
Canto, J., $\&$  Raga, A. C. 1998, Nature, 395, 355
\bibitem[Th\'ebault et al. (2004)]{Thebault04}
Th\'ebault, P., Marzari, F., Scholl, H., Turrini, D., \& Barbieri, M.
2004, \aa, 427, 1097
\bibitem[Th\'ebault et al. (2006)]{Thebault06}
Th\'ebault, P., Marzari, F., \& Scholl, H., 2006, 
Icarus, 183, 193
\bibitem[Turrini et al. (2005)]{Turrini05}
Turrini, D., Barbieri, M., Marzari, F., Th\'ebault, P., $\&$ Tricarico, P.
2005, Memor.Soc.Astron.It.Suppl., 6, 172
\bibitem[Turrini et al. (2006)]{Turrini06}
Turrini, D., Barbieri, M., Marzari, F., Th\'ebault, P., $\&$ Tricarico, P.
2005, Memor.Soc.Astron.It.Suppl., 9, 186
\bibitem[Weidenschilling (1977)]{Stu77}
Weidenschilling, S. J. 1977, Astrophysics \& Space Science, 51, 153
\bibitem[Wetherill (1996)]{Wetherill96}
Wetherill, G. W. 1996, Icarus, 119, 219
\bibitem[Whitmire et al. (1998)]{Whitmire98}
Whitmire, D. P., Matese, J., L., Criswell, L., \&
Mikkola, S. 1998, Icarus, 132, 196
\end{thebibliography}
\end{document}